\documentclass[pra,twocolumn,nofootinbib]{revtex4}

\usepackage{graphicx}
\usepackage{dcolumn}
\usepackage{bm}
\usepackage{amssymb}
\usepackage{amsmath}
\usepackage{amsfonts}
\usepackage[]{graphicx}
\def\beq{\begin{eqnarray}}
\def\eeq{\end{eqnarray}}

\begin{document}
\bibliographystyle{prsty}

\title{The Quantum  Eraser Non-Paradox: a comment on arXiv:2111.09347v1}
\author{Aur\'elien Drezet}
\affiliation{Institut N\'eel, UPR 2940, CNRS-Universit\'e Joseph Fourier, 25, rue des Martyrs, 38000 Grenoble, France}

\date{\today}

\begin{abstract}
This is a short reply to arXiv:2112.00436v1 and arXiv:2111.09347v1 by C. Bracken,  J.R. Hance, and S. Hossenfelder. I precise some elements concerning i) the methodology of my previous version of the comment [arXiv:2111.13357v3] to arXiv:2111.09347v1, and ii) why the grand father-paradox (causal loop) presented in their article comes from a prejudice concerning retrocausality not from retrocausality or locality themselves.  
\end{abstract}
 \maketitle
\section{The rebuttal}
\indent In their reply \cite{Sabine2} to my comment \cite{Drezet} of their `quantum eraser paradox' \cite{Sabine} C. Bracken,  J.R. Hance, and S. Hossenfelder claim that my analysis of their causal loop paradox   fails to refute their conclusions.  However, their answer \cite{Sabine2}, like  \cite{Sabine}, is still very confusing. In \cite{Sabine} the authors propose a kind a contingent detection protocol in order to produce a causal loop paradox.  One of the key ambiguity in \cite{Sabine} that I tried to clarify with the authors was the meaning of `turning on detector $D_1$' (\cite{Sabine}, page 4) induced by a detection at $U_4$. In their proposal there is no switchable mirror  but a passive beam splitter and turning on seems to imply that they only make a sampling  or filtering on the data. Indeed, a photon could go into the arm $D_1$ even if the detector is off (it could then be absorbed by a shutter). This represents a loss.  It would thus be unnecessary to turn on the detector: i.e., just collecting all the data and then filtering out  the unwanted coincidences $D_1,U_3$ would be equivalent. It is not clear what missing some data implies on the whole deduction~\footnote{ I mention \textit{en passant} that debunking the original Sutherland paradox \cite{Sutherland} is easily done after realizing that neglecting (discarding) half of the runs leads to a contradiction.  }.  Also in their proposal there is no detector $D_2$ and no associated beam splitter. This makes the intensities arriving on $D_3,D_4$  asymmetric and the analysis will thus be more complicated. Their setup is therefore very confusing  and to debunk the paradox it is  necessary 1) to clearly define the `contingent-choice' protocol by removing ambiguities and 2) to provide a calculation of the standard quantum predictions. All this was not present in \cite{Sabine}.\\        
\indent In \cite{Drezet} the aim was twofold: 1) to define such a non ambiguous ` contingent-detection' protocol (see my Figure 1 in \cite{Drezet}) without the previous potential problems (not analyzed in \cite{Sabine}), and 2) to  show that a clean analysis of the contingent measurement protocol requires  entanglement with the full device. It is so because the feedback loop involved in the process relies on the quantum correlation between what are occurring in gate $U_4$ and gate $D_1$.  The final state is a macroscopic entanglement involving pointers and as I wrote decoherence is central. It means that it doesn't really matter if we cannot act on this giant   `Schr\"odinger cat' state to see some interference effects at the macroscale (as I mentionned this is related to the Wigner friend's paradox). This answers to their recent claim that `no one has ever observed such thing'~\cite{Sabine2} (we also discussed this issue through emails).  For all practical purposes  (FAPP\footnote{Of course, I never asumed that the outcomes or recorded eigenstates are not well defined in the end: I am myself a strong advocate of the de Broglie-Bohm theory and FAPP is not enough. We need an ontology. Moreover, in Bohm's theory we have a lot of empty waves and only a full unitary calculation can define the particle dynamics guided by the waves.} ) decoherence defines a preferred basis and the quantum state I obtained (Eq. 10 in \cite{Drezet})  is necessarily the good one to use in order to describe the quantum measurement and obtain the quantum probabilities. Since the authors of \cite{Sabine} didn't provide any details on the wave functions in their setup it is rather impossible to make a clear deduction. The authors of \cite{Sabine,Sabine2} believe that classical detectors are enough to understand their protocol. In a Copenhagen reading this is indeed  true.  But again one has to be careful. As I stressed in \cite{Drezet} in the standard quantum measurement theory the position of the quantum-classical boundary is irrelevant as far as we dont create paradoxes similar to the one proposed by Wigner. This motivated the calculations made in \cite{Drezet} to see if it is so.  Moreover, this is also central because the authors of \cite{Sabine} seem to rely on a idea of collapse which creates a time-symmetry breaking or an irreversibility (they wrote in \cite{Sabine}: `Note, in this local-realist setting, we assume measurement at the detector collapses the wave function'). However, in most retrocausal models we need a kind of time symmetry involving two waves (one going forward in time the second backward).   The concept of collapse  coupled to retrocausal hidden-variables (that they also mention as one of their target) is therefore problematic.\\ 
\indent  It is not difficult to see why this is problematic and we  need neither switching device nor contingent detection to prove it.  Penrose already emphasized this issue~\cite{Penrose}.      Consider for example a single photon on a symmetric beam splitter: $|\gamma_0\rangle\rightarrow\frac{1}{\sqrt{2}}|\gamma_t\rangle +i\frac{1}{\sqrt{2}}|\gamma_r\rangle$. If we assume a collapse in the exit gate `r' we can watch the problem backward in time and see that we have now  $\frac{1}{\sqrt{2}}|\gamma_1\rangle -i\frac{1}{\sqrt{2}}|\gamma_0\rangle\leftarrow|\gamma_r\rangle$. This is of course looking weird due to the presence of the state $|\gamma_1\rangle$ in the second entrance gate which is supposed to be empty: uncollapsing is problematic. This could lead to physical contradictions. For instance, consider now an EPR superposition like $\frac{1}{\sqrt{2}}(|\gamma_0\rangle|\Gamma_0\rangle+|\gamma_1\rangle|\Gamma_1\rangle)$  where $\Gamma_{0,1}$ are orthogonal which path qubit states  for the two previous photon states  $|\gamma_{0,1}\rangle$.  If we let the two photon states interacting with the previous beam splitter we end up with:
\begin{eqnarray}
\frac{1}{\sqrt{2}}(|\gamma_0\rangle|\Gamma_0\rangle+|\gamma_1\rangle|\Gamma_1\rangle)\rightarrow \nonumber\\
\frac{1}{\sqrt{2}}[|\gamma_t\rangle(\frac{|\Gamma_0\rangle+i|\Gamma_1\rangle}{\sqrt{2}})+|\gamma_r\rangle(\frac{|\Gamma_1\rangle+i|\Gamma_0\rangle}{\sqrt{2}})]. 
\end{eqnarray} Now, suppose a collapse at exit gates $\Gamma_0$ and  $\gamma_r$. If we watch the process backward in time we have:
\begin{eqnarray}
\frac{1}{\sqrt{2}}(|\gamma_1\rangle|\Gamma_0\rangle-i|\gamma_0\rangle|\Gamma_0\rangle)\leftarrow|\gamma_r\rangle|\Gamma_0\rangle.
\end{eqnarray} This is clearly paradoxical if we try to interpret the term $|\gamma_1\rangle|\Gamma_0\rangle$:  A photon came back to the `0' entrance gate but at the same time we have the $\Gamma_1$ which-path state that seems to imply that the photon went to  the `0' gate. Moreover, if it actually went to the `0' gate then should we have the $\Gamma_0$ or the $\Gamma_1$ state? In fact if we watch the problem even more backward in time we could realize that probably the creation of a pair $|\gamma_1\rangle|\Gamma_0\rangle$ is forbidden by some laws of conservations (this could be so if the which-path qubit is also a photon and if the pair  of photons  is generated by a local parametric fluorescence process). Therefore, there is a physical contradiction.  As explained by Penrose: `Of course the [collapse] rule was not designed to be applied into the past, but it is instructive  to see how completely wrong it would be to be so'~\cite{Penrose}. I believe that the same kind of problem occurs in the proposal~\cite{Sabine}: The problem is with the collapse-rule not (necessarily) with retrocausality. Their proposal is not sufficient to refute retrocausality  but perhaps only a small subset of such models. These worries explain also why I used a quantum description of the detection protocol (e.g., this could be used to give a complete description of the paradox using the Sutherland-Sen retrocausal hidden-variable model without collapse).\\
\indent Going back to our problem: as I explained in \cite{Drezet} the authors of  \cite{Sabine} seem to believe that in a retrocausal and local model a green photon detected in the gate $D_1$ could send back in time a message to the point B of my Figure 1 in \cite{Drezet} then an information could go forwardly in time  with the other red photon to force him to go to gate $U_3$ even though this would lead to a contradiction when  we use the switching device. But why do they think so?  Again probably because of the idea of collapse I discussed before or perhaps because they mixed different configurations and experimental setups as explained  in \cite{Drezet}. Indeed, with fixed mirrors we have a which-path experiment and no interference between beams going to $U_3$ and $U_4$. We could thus naively think that half of the runs should go to the wrong gate, i.e., $U_3$ even though this is impossible with the switching device (that was the aim of my calculations in \cite{Drezet} to fix this issue).   In my opinion, this only shows some prejudices of the authors concerning retrocausal models but doesn't prove that the red photon should behave as they claim.\\  
\indent This is like the old grand father paradox:  How to avoid this?   Easy:  when you are going back in the past you just cannot kill your grand father.  Why you cannot? Because there is only one Universe   you can not create paradox. May be something will prohibit you to kill the grand-father. You will be killed by a car or something like that. In the end free-will is an illusion in a deterministic Universe and it is even more important to remember this point with retrocausal and superdeterministic approaches.\\
\indent In the example of \cite{Sabine} the photon has no free-will and the information cannot be used to force the red photon to go to $U_3$ this would lead to a contradiction: End of the paradox. You don't have to suppose that perhaps `the world end in a poof'~\cite{Sabine}.  Note that locality is not at stake or in danger here.  We can perfectly well assume locality and still avoid the paradox. The two photons have enough  shared informations (with the switching time-like or light-signal going from $U_{3,4}$ to $D_{1,2,3,4}$) to avoid the paradox. If a green photon is detected in $D_1$  the detector also receives a signal from the detector $U_4$ and it can send an information backward in time to B in order that  the other red phtoton is going to $U_4$ but never to $U_3$. The causal loop is avoided: Retrocausality cannot be dismissed so easily!


\end{document}